\documentclass[11pt,letterpaper]{article}

\usepackage{geometry}
\usepackage{bbm}
\usepackage[toc,page]{appendix}
\usepackage{amsmath, amsthm, amsfonts, amssymb}
\usepackage{mathrsfs} 
\usepackage{hyperref}
\usepackage{cleveref}
\usepackage{slashed}
\usepackage{dsfont}
\usepackage{wrapfig}
\usepackage{minibox}
\usepackage{graphicx}
\usepackage{graphics}
\usepackage{epsfig}
\usepackage{xcolor}
\usepackage{array}
\usepackage{tikz}
\usepackage{tikz-cd}
\usetikzlibrary{arrows,calc}
\usepackage{relsize}
\usetikzlibrary{decorations.pathmorphing,calc}
\usepackage{authblk}
\usepackage[labelfont=bf]{caption}
\usepackage{caption}
\usepackage{subcaption}
\usepackage{placeins}
\usepackage{braket}

\providecommand{\href}[2]{#2}

\def\a{\alpha}
\def\b{\beta}
\def\g{\gamma}
\def\G{\Gamma}
\def\d{\delta}
\def\D{\Delta}
\def\ve{\varepsilon}
\def\m{\mu}
\def\n{\nu}

\def\L{\Lambda}
\def\k{\kappa}
\def\r{\rho}
\def\s{\sigma}
\def\thintablerule{\hrule height0.4pt}

\def\bm{\bold}

\newcommand{\be}{\begin{equation}}
\newcommand{\ee}{\end{equation}}
\newcommand{\bea}{\begin{eqnarray}}
\newcommand{\eea}{\end{eqnarray}}
\newcommand{\bal}{\begin{aligned}}
\newcommand{\eal}{\end{aligned}}
\newcommand{\eq}[1]{Eq.~(\ref{#1})}

\newcommand{\sect}[1]{Section~\ref{#1}}

\numberwithin{equation}{section}

\newcommand{\nbad}{\begin{equation*} \begin{aligned}}
\newcommand{\nead}{\end{aligned} \end{equation*} }
\newcommand{\bad}{\begin{equation} \begin{aligned}}
\newcommand{\ead}{\end{aligned} \end{equation}}
\numberwithin{equation}{section}
\newcommand{\nl}{\nonumber \\}

\def\thintablerule{\hrule height0.4pt}

\def\bm{\bold}

\def\tk{\bm k}
\def\tmk{{|\bm k|}}

\def\tq{\bm q}

\def\tx{{\bm x}}

\def\ty{{\bm y}}

\def\o{\omega}
\def\O{\Omega}

\allowdisplaybreaks

\begin{document}

\tikzset{
    photon/.style={decorate, decoration={snake}, draw=black},
    electron/.style={draw=black, postaction={decorate},
        decoration={markings,mark=at position .55 with {\arrow[draw=black]{>}}}},
    gluon/.style={decorate, draw=black,
        decoration={coil,amplitude=4pt, segment length=5pt}} 
}

\centerline{\Huge Thermal effects in Ising Cosmology}

\vskip 1 cm
\centerline{\large Nikos Irges$^{\spadesuit}$\footnote{e-mail: irges@mail.ntua.gr}, Antonis Kalogirou$^{\spadesuit}$\footnote{e-mail: akalogirou@mail.ntua.gr}
and Fotis Koutroulis$^{\clubsuit}$\footnote{e-mail: fotis.koutroulis@fuw.edu.pl}}
\vskip 1cm
\centerline{\it $\spadesuit$. Department of Physics}
\centerline{\it School of Applied Mathematical and Physical Sciences}
\centerline{\it National Technical University of Athens}
\centerline{\it Zografou Campus, GR-15780 Athens, Greece}
\vskip .5cm
\centerline{\it $\clubsuit$. Institute of Theoretical Physics, Faculty of Physics,}
\centerline{\it University of Warsaw, Pasteura 5, PL 02-093, Warsaw, Poland}

\vskip 2.2 true cm
\thintablerule
\vskip 2.0ex

\centerline{\bf Abstract}
We consider a real scalar field in de Sitter background and compute its thermal propagators.
We propose that in a dS/CFT context, non-trivial thermal effects as seen by an `out' observer can be encoded in the anomalous dimensions of the $d=3$ Ising model.
One of these anomalous dimensions, the critical exponent $\eta$, fixes completely a number of cosmological observables, which we compute.

\vskip 1.0ex\noindent
\vskip 2.0ex
\thintablerule

\newpage

\pagebreak


\section{Introduction}

The rapidly expanding phase of the universe can be modelled by de Sitter (dS) space
and the simplest form of matter by a real scalar. It is believed that basic effects that left an imprint
on the Cosmic Microwave Background (CMB) were of thermal nature.
Therefore a simple model that could explain some of the observed features of the CMB is a real scalar field $\phi$ in the expanding
Poincare patch of dS space \cite{BirrellDavies,Mukhanov,Akhmedov1}, formulated in the context of thermal quantum field theory \cite{Das2}. 
The action is 
\be
{ \cal S} = \int  d^4 x  \ \sqrt{-g} \left[ \frac{1}{2} g^{\mu \nu} \partial_\mu \phi \partial_\nu \phi - \frac{1}{2}(m^2 + \xi {\cal R}) \phi^2 \right]\, ,
\ee
which we will quantize taking into account finite temperature effects. Here $m$ is the mass of the scalar field, $\xi$ is the non-minimal coupling of the field to gravity and $\cal R$ the scalar curvature. 

Consider a $d+1$ dimensional FRW spacetime with metric
\be
ds^2 = a^2 \left(d \tau^2 - d  \bm x^2 \right)
\ee
with $\tau $ the conformal time and $a(\tau )$ the scale factor. The expanding Poincare patch of de Sitter space corresponds to 
$a=-\frac{1}{H \tau }$ where $H =  \frac{a'}{a^2}$ the Hubble constant with $a' = \frac{d a}{d \tau}$. The expanding Poincare patch of dS space is parametrized by $\tau \in (-\infty, 0]$.
The scalar field mode in d-dimensional momentum space $\phi_\tmk = \frac{\chi_\tmk}{a}$ in this background yields the classical Klein-Gordon equation of motion
($k=(k^0, \tk)$ is the four-momentum and the prime is derivative with respect to $\tau $)
\be\label{eom}
{ \chi_{\tk}^{\prime\prime}} + \omega^2_\tmk \chi_{\tk} = 0\, ,
\ee
with $ \o^2_\tmk  = \tmk^2  + m_{\rm dS}^2 $ and a time-dependent mass given by $m_{\rm dS}^2=\frac{1}{\tau ^2} (M^2 - \frac{d^2 -1}{4})$.
The dS mass parameter is $M^2=\m_H^2+12\xi$ with $\m_H^2=\frac{m^2}{H^2}$ and $H$ the inverse curvature parameter of dS space, satisfying ${\cal R}=12 H^2$.
The solutions to \eq{eom} are linear combinations of the Hankel function $H_{\n_{\rm cl}}(\tau ,\tmk)$ and its complex conjugate, of weight $\n_{\rm cl}$, with
\be\label{ncl}
\n_{\rm cl} = \frac{d}{2} \sqrt{1 - \frac{4 M^2}{d^2} }
\ee
and as we will revisit later on, $\n_{\rm cl}$ is a part of the classical scaling dimensions of the bulk and boundary operators. 

Focusing on a given vacum at a specific time, we can Taylor expand in the usual way the scalar field and calculate the Hamiltonian \cite{Mukhanov}:
\bad \label{Hamilt}
{\cal H} = \frac{1}{4} \int d^3 k \left[ \Omega_\tmk \left( 2 \a^\dagger_\tk \a_\tk + \delta^{(3)} (0) \right) + \Lambda_\tmk \a_\tk^\dagger \a_{-\tk}^\dagger + \Lambda^*_\tmk \a_\tk \a_{-\tk} \right] 
\ead 
where $\a_\tk, \a_\tk^\dagger$ are the annihilation and creation operators that satisfy the commutation relations:
\bad
\Bigl[\a_\tk, \a_\tq^\dagger \Bigr] = \delta (\tk - \tq), \qquad \Bigl[\a_\tk, \a_\tq \Bigr] = \Bigl[\a_\tk^\dagger, \a_\tq^\dagger \Bigr] =0.
\ead
In addition, $\Omega_\tmk$ and $\Lambda_\tmk$ are defined as
\bad
\Omega_\tmk = | u'_\tmk |^2 + \omega^2_\tmk |u_\tmk |^2, \qquad \Lambda_\tmk =  {u'}^2_\tmk + \omega^2_\tmk u_\tmk^2
\ead
where $u_\tmk,u^*_\tmk$ are the mode functions that pair with the ladder operators $\a_\tk,\a^\dagger_\tmk$. 
The frequency $\omega^2_\tmk$ will be defined below and it carries the time dependence.
One of the main key points of a QFT in a curved spacetime is that there is not a single choice for a vacuum state, while different choices lead to different ladder operators 
$\b_\tk, \b^\dagger_\tk$ and mode functions that are connected by the Bogolyubov Transformation (BT) \cite{Mukhanov}:
\bad
\b_\tk = c_\tmk \a_\tk + d^*_\tmk \a_{- \tk}^\dagger, \qquad \b_\tk^\dagger = c^*_\tmk \a_\tk^\dagger + d_\tmk \a_{-\tk}
\ead
with $c_\tmk, d_\tmk$ the Bogolyubov coefficients. Using the commutation relations defined above, one can show in a straightforward manner 
that the Hamiltonian \eqref{Hamilt} is hermitian for a given ground state defined at a given time $\tau$.\footnote{An analysis concerning the full time range that spans the entire dS space
(instead of only its expanding patch that we consider here) is more delicate and Hermiticity maybe lost.}

The work presented in this paper follows the related work \cite{FotisAntonis1} where a possible connection between a bulk dS theory 
and a boundary Ising model was examined along with its implications to the value of the cosmological spectral index $n_S$. In particular, 
experiments \cite{Planck2018X},\cite{Planck2018IX} find that $n_S$ deviates slightly from unity which shows that the CMB is nearly scale invariant. 
Although, it is true that other endeavors to explain this deviation exist (e.g. \cite{Maldacena1,Larsen}), the idea of using the critical exponent 
of a boundary Ising field in order to predict cosmological observables is new. Furthermore, it is important to note that the calculations 
and results of the current paper are model-independent, meaning that we only take for granted the experimental value of $n_S$ and the 
assumption that the inflation era of the Universe can be explained by the expanding Poincare patch of dS spacetime.

In \sect{PropandTemp} we compute thermal propagators in dS spacetime by generalizing methods first produced in flat spacetime 
(as firmly discussed in \cite{Das2}) i.e. the Schwinger-Keldysh (SK) path integral and the Thermofield Dynamics (TFD). To our knowledge, 
while there are cases upon which the SK path integral has been used in dS before (e.g. \cite{Chen}), the use of the TFD formalism in a 
general FRW spacetime is novel. Moreover, a plain but non-trivial connection between the two formalisms is presented which can hold 
for other spacetime choices other than dS and enables one to avoid ambiguities rising in the usual SK construction. In \sect{SpectralIndex}, 
we make use of the dS thermal propagators in order to incorporate the thermal corrections that arise when one proceeds to calculate the scalar 
spectral index $n_S$. In contrast to previous works \cite{MolinaParis} where the spectral index to leading order was found to be equal to unity, here we argue
that thermal effects actually slightly break scale invariance, resulting in $n_S \neq 1$. This leads to a parametric freedom which can be fixed by 
an RG flow argument that has its origins in the $d=3$ Ising model and in such a way, we can match the deviation of $n_S$ away from unity to the
experimental data. Finally, using the newly formed thermal dS propagator, we can extract more cosmological observables determined by the scalar metric fluctuations, 
namely the running of the spectral index $n^{(1)}_S$ and the non-Gaussianity parameter $f_{NL}$.

\section{Propagators and temperature} \label{PropandTemp}

Quantization of this system results in the notion of a time-dependent vacuum state and a doubled Hilbert space.
Regarding the vacua, we will be concerned with the so called ``in" vacuum defined at $\tau =-\infty$
and the ``out" vacuum defined at the boundary (i.e. the horizon) of the expanding patch, at $\tau =0$. 
These are empty vacua from the perspective of corresponding local (in conformal time) observers. 
The $\ket{\rm in}$ will be chosen to be the maximally symmetric Bunch-Davies (BD) vacuum \cite{Chernikov,Allen}.
The two vacua are related via the BT 
$\bra{J}{\bf\Phi}^{I} = \bra{I}{\bf\Phi}^{J}$ where $I,J = {\rm in}, {\rm out}$ is a label of the vacuum and ${\bf\Phi}^I$ is the field operator
with mode function $\chi^I_\tmk$. Note that the field is the same in both vacua, with the mode functions and the creation and annihilation operators
inside it being the vacuum dependent quantities.
Common notation is $\chi^{\rm in}_\tmk=u_\tmk$ and $\chi^{\rm out}_\tmk=v_\tmk$.

The doubled Hilbert space can be understood in the context of the SK path integral as being related to a 
$+$ (or forward) branch and a $-$ (or backward) branch in conformal time evolution. 
The field propagator ${\cal D}$ in such a basis has a $2\times 2$ matrix structure and is (${\cal T}$ (${\cal T}^*$) 
denoting time (anti-time) ordering and $\bra{0}$ is a generic vacuum):
\bea\label{SKzTSm}
\bra{0}  \Phi^{+}(\tau _2) \Phi^{-}(\tau _1) \ket{0} &=& {\cal D}_{-+}(\tau _1; \tau _2) \nonumber\\
\bra{0} \Phi^{-}(\tau _1)  \Phi^{+}(\tau _2) \ket{0} &=& {\cal D}_{+-}(\tau _1; \tau _2)
\eea
and
\bea\label{SKzTSm2}
\bra{0}{\cal T} [  \Phi^{+}(\tau _1)  \Phi^{+}(\tau _2) ] \ket{0}  &=& {\cal D}_{++}(\tau _1;\tau _2) \nonumber\\
\bra{0} {\cal T}^* [\Phi^{-}(\tau _1) \Phi^{-}(\tau _2) ] \ket{0} &=& {\cal D}_{--}(\tau _1; \tau _2)
\eea
where ${\cal D}_{+-}(\tau _1; \tau _2) = {\cal D}^{*}_{-+}(\tau _1; \tau _2)$, ${\cal D}_{--}(\tau _1; \tau _2) = {\cal D}^{*}_{++}(\tau _1; \tau _2)$
and
${\cal D}_{-+}(\tau _1; \tau _2) = \chi_{\tmk}(\tau _1) \chi^{*}_{\tmk}(\tau _2)$, 
${\cal D}_{++}(\tau _1;\tau _2)  = \theta( \tau _1 - \tau _2 ) {\cal D}_{-+}(\tau _1; \tau _2) +  \theta( \tau _2 - \tau _1 ) {\cal D}_{+-}(\tau _1; \tau _2) $. Note that, since  the vacuum state $\ket{0}$ is time dependent, we construct the above formulas without choosing a specific vacuum for now. In addition, there is no need for a time ordered product in \eq{SKzTSm} because the two fields commute since they are defined in different parts of the SK contour. 

The above matrix elements satisfy the relation 
\bad \label{KMSSK}
{\cal D}_{++} + {\cal D}_{--} - {\cal D}_{+-} - {\cal D}_{-+} = 0.
\ead 
Hidden in these expressions is the $i\ve$ shift, implementing the projection on the vacuum at $\tau =-\infty$.
It can be chosen so that in the flat limit the propagator becomes diagonal with ${\cal D}_{++}=\frac{-i}{k^2-m^2+i\ve}$. 
The above construction of the propagator at zero temperature in dS spacetime has been recently studied in \cite{Chen}.

The thermal generalization of the propagator components in \eq{SKzTSm} and \eq{SKzTSm2} is our next goal.
If the Hamiltonian of the system was time-independent, one could just follow the process described in Appendix \ref{AppKMS}
and show that the propagator satisfies the KMS condition \cite{KMS}, which ensures that it is a good thermal propagator.
Here however we are dealing with a time-dependent Hamiltonian and this is not straightforward. 
Instead, we will use the method introduced in \cite{Semenoff} that takes advantage of the SK contour,
by adding an extra,``thermal" leg to it. In particular, if ${\cal C}_+$ is the forward branch 
where time evolution follows the path $\tau_{\rm in} \rightarrow \tau_{\rm out}$, ${\cal C}_-$ is the backward branch where 
$\tau_{\rm out} \rightarrow \tau_{\rm in}$, we attach an extra part to the contour ${\cal C}_3$, where $\tau_{\rm in} \rightarrow \tau_{\rm in} - i \frac{\beta}{2}$
and $\beta=1/T$ is the inverse temperature parameter:
%
\begin{figure}[!htbp]
\centering
\includegraphics[width=6cm]{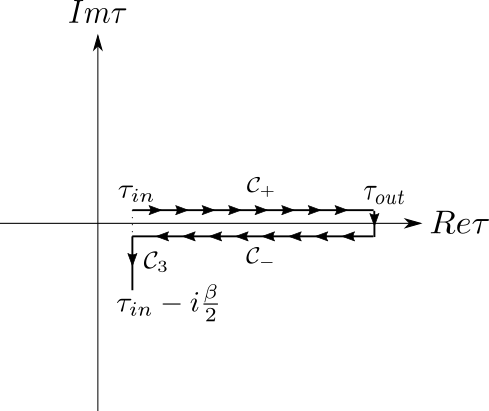}
\end{figure}\label{ThermalSK}
\FloatBarrier
%
Furthermore, we introduce the propagators
\bea
&& \bra{0}{\cal T} [  \Phi^{3}(\tau _1)  \Phi^{3}(\tau_2) ] \ket{0}  = {\cal D}_{33}(\tau_1;\tau_2)   \nonumber \\
&& \bra{0}  \Phi^{+}(\tau_1) \Phi^{3}(\tau_2) \ket{0} = {\cal D}_{3+} (\tau_1; \tau_2) \nonumber \\
&& \bra{0}  \Phi^{-}(\tau _1) \Phi^{3}(\tau_2) \ket{0} = {\cal D}_{3-} (\tau_1; \tau_2)
\eea
where $\Phi^{3}(\tau)$ is the field operator living on ${\cal C}_3$ and $\tau_1,\tau_2 \in {\mathbb C}$ and we demand that the junction conditions for $a  \in \{+,-,3 \}$: 
\begin{gather}
{\cal D}_{a+} (\tau_1 ; \tau_2 ) \biggl|_{\tau_2 = \tau_{\rm out}}  = {\cal D}_{a-} (\tau_1 ; \tau_2 ) \biggl|_{\tau_2 = 
\tau_{\rm out}} \qquad \frac{\partial}{\partial \tau_2} {\cal D}_{a+} (\tau_1 ; \tau_2 ) \biggl|_{\tau_2 = \tau_{\rm out}}  = 
\frac{\partial}{\partial \tau_2} {\cal D}_{a-} (\tau_1 ; \tau_2 ) \biggl|_{\tau_2 = \tau_{\rm out}} 
\end{gather}
are satisfied at the time instance $\tau = \tau_{\rm out}$ where the ${\cal C}_+$ and ${\cal C}_-$ contours meet, while the conditions
\begin{gather}
{\cal D}_{a-} (\tau_1 ; \tau_2 ) \biggl|_{\tau_2 = \tau_{\rm in}}  = {\cal D}_{a3} (\tau_1 ; \tau_2 ) \biggl|_{\tau_2 = \tau_{\rm in}} \qquad 
\frac{\partial}{\partial \tau_2} {\cal D}_{a-} (\tau_1 ; \tau_2 ) \biggl|_{\tau_2 = \tau_{\rm in}}  = \frac{\partial}{\partial \tau_2} {\cal D}_{a3} (\tau_1 ; \tau_2 ) \biggl|_{\tau_2 = \tau_{\rm in}} 
\end{gather}
 need to be satisfied at $\tau = \tau_{\rm in}$ where ${\cal C}_-$ and ${\cal C}_3$ meet. Finally for the SK analogue of the KMS condition to hold, 
 we need to sew together ${\cal C}_+$ and ${\cal C}_3$ which results in the conditions
\bad
{\cal D}_{a+} (\tau_1 ; \tau_2 ) \biggl|_{\tau_2 = \tau_{\rm in}}  = {\cal D}_{a3} (\tau_1 ; \tau_2 ) \biggl|_{\tau_2 = \tau_{in} - i \beta/2} \qquad 
\frac{\partial}{\partial \tau_2} {\cal D}_{a+} (\tau_1 ; \tau_2 ) \biggl|_{\tau_2 = \tau_{\rm in}}  = \frac{\partial}{\partial \tau_2} {\cal D}_{a3} (\tau_1 ; \tau_2 ) \biggl|_{\tau_2 =\tau_{\rm in} - i  \beta/2}
\ead
that ensure the consistency of the deformed contour and yield a good thermal propagator.

The above conditions will introduce corrections of thermal nature into the propagators \eq{SKzTSm} and \eqref{SKzTSm2},
which we compute by making two assumptions. Since the chosen contour allows for an imaginary time flow, 
we assume that there is no inflation in that direction. This means that the mode functions living on the ${\cal C}_3$ 
leg of the contour can be taken to have a plane wave form. In addition, at $\tau = \tau_{in}$ we assume the BD vacuum 
so that the mode functions are expressed in terms of the Hankel functions of $\nu_{\rm cl} = 3/2$ order.   
According to these assumptions, the solution to the conditions results in the in-in thermal propagator components \cite{Semenoff}:
\begin{align}
{\cal D}^{\beta/2}_{++}  &= {\cal D}_{++} + n_B(\beta/2) \left({\cal D}_{++}  + {\cal D}_{--} \right) \nonumber \\ 
{\cal D}^{\beta/2}_{--}  &= {\cal D}_{--} + n_B(\beta/2) \left({\cal D}_{++}  + {\cal D}_{--} \right) \nonumber \\ 
{\cal D}^{\beta/2}_{+-}  &= {\cal D}_{+-} + n_B(\beta/2) \left({\cal D}_{++}  + {\cal D}_{--} \right) \nonumber \\ 
{\cal D}^{\beta/2}_{-+}  &= {\cal D}_{-+} + n_B(\beta/2) \left({\cal D}_{++}  + {\cal D}_{--} \right) 
\end{align}
with $n_B$ the Bose-Einstein distribution parameter
\be
n_B(\beta) =  \frac{e^{-\b \o_\tmk}}{{1-e^{-\b \o_\tmk}}}\, .
\ee
We can express conveniently this propagator collectively in a matrix notation as:
\bad \label{SKDininbeta}
{\cal D}_{\beta/2} = {\cal D} + s^2(\beta/2) \left({\cal D}_{++}  + {\cal D}^*_{++} \right) 
\begin{pmatrix}
1 & 1 \\
1& 1
\end{pmatrix}
\ead    
with
\nbad
{\cal D} = \begin{pmatrix}
{\cal D}_{++} & {\cal D}_{+-} \\
{\cal D}_{-+} & {\cal D}_{--} \\
\end{pmatrix}, 
\nead
and the parametrization $s(\beta/2) \equiv \sinh \theta_\tmk(\beta/2) = \sqrt{n_B(\beta/2)}$ and $c(\beta/2)\equiv\cosh\theta_\tmk(\beta/2)$.\footnote{The flat limit of this propagator is diagonal and its
$++$ component is such that the $i\ve$ shift of the zero temperature propagator
denominator becomes $iE = i \ve \coth(\b\o_\tmk/2)$ in the thermal state.}
It is easy to see that this thermal propagator satisfies a condition like \eq{KMSSK}. 

Here we are actually interested in the out-out thermal propagator.
We will first derive the result using a novel shortcut and then we will show that it indeed yields the correct result.
The shortcut uses the TFD formalism, where the doubled Hilbert space is
seen as the tensor product of the Hilbert spaces of positive and negative momenta ${\cal H}$ and ${\tilde{\cal H}}$.
The fields living in these Hilbert spaces are $\Phi$ and ${\tilde\Phi}$ correspondingly.
The validity of this strategy is based on the fact that the SK structure can be read also as a TFD structure, 
in which case the passage to finite temperature is via the transformation ${\cal D}_ {\b'} = U_{\b'}\, {\cal D}\, U^{T}_{\b'}$ and \cite{Takahashi}
\be\label{unitaryDas}
U_{\b'} \equiv \begin{pmatrix}
\cosh \theta_\tmk(\beta') & \sinh \theta_\tmk(\beta')  \\
\sinh \theta_\tmk(\beta')  & \cosh \theta_\tmk(\beta')
\end{pmatrix}\, .
\ee
That this is an allowed operation on dS propagators is supported by the fact that a transformation by the matrix $U_{\b'}$ is a BT with coefficients
$\sinh \theta_\tmk(\beta') = \frac{e^{-\frac{\b'}{2} \o_\tmk}}{\sqrt{1-e^{-\b' \o_\tmk}}}$ and $\cosh \theta_\tmk(\beta') = \frac{1}{\sqrt{1-e^{-\b' \o_\tmk}}}$.
Hence, we essentially calculate the thermal corrections that the BT has on the propagator via the TFD formalism.
The result of the rotation gives the out-out thermal propagator
\be\label{Doutoutb}
{\cal D}_{\b'}  =  {\cal D} + (s^2(\beta')+s(\beta')c(\beta')) \left({\cal D}_{++} + {\cal D}_{++}^ {*} \right) 
 \begin{pmatrix}  1 & 1 \\ 1 & 1 \end{pmatrix} \, .
\ee
One  immediately notices that the two expressions in \eq{SKDininbeta} and \eq{Doutoutb} disagree
in the thermal correction, as the latter has an extra term along $\sinh \theta_\tmk \cosh \theta_\tmk$. 
This might seem troublesome at first, however they both contain the same physical information.
Taking advantage of the trivial identity
\bad
\frac{e^{- \beta \o_\tmk}}{1 - e^{- \beta \o_\tmk}} + \frac{e^{- \frac{\beta}{2} \o_\tmk}}{1 - e^{- \beta \o_\tmk}} = \frac{e^{- \frac{\beta }{2}\o_\tmk}}{1 - e^{- \frac{\beta }{2}\o_\tmk}}\, ,
\ead
the propagators in \eq{SKDininbeta} and \eq{Doutoutb} are seen to be equal for $\beta' = \beta$. 
Note that the above identity does hold in the $\sinh \theta_\tmk$ and $\cosh \theta_\tmk$ parametrization, where it reads
$s^2(\beta)+s(\beta)c(\beta) = s^2(\beta/2)$.
We have therefore proved that the known form of the dS thermal propagator of \cite{Semenoff} can be equivalently obtained via a TFD rotation
of the zero temperature SK propagator of the half thermal parameter. 
The equivalence of the two expressions reflects of course the universal nature of the dS temperature as measured at an arbitrary time instance by
the in and out observers. The advantage of the TFD rotation operation is that it is very simple and can be 
easily generalized to any background. Thus, we will use this point of view in the following.

The result of all allowed thermal transformations of ${\cal D}$ are correlators of the form
\be\label{4props}
{\cal D}^I_ {J,\g}
= \bra{{J};\g}  {\cal T} [ {\bf\Phi}^{I} ({\bf\Phi}^{I})^T ] \ket{{J};\g} \, .
\ee
The doublet field, now in the language of TFD,
is $({\bf\Phi}^{I})^T = ( \Phi^{I}, {\tilde\Phi}^{I})$ and $\g$ is a thermal index, associated with any combination of thermal transformations of the form \eq{unitaryDas}.
The label (not index) $I$ on the field is a reminder of the vacuum state to which the mode functions belong.
The two types of thermal transformations that are relevant to us are the insertion of an explicit density matrix, resulting in a transformation by a unitary operator $U$, as
$\ket{I;\b} =  {U} \ket{ I}$, where the eigenvalue of $U$ is $U_\b(\theta)$ and the Gibbons-Hawking (GH) effect \cite{GibbonsHawking} (for which we will momentarily use the parameter $\d$
to distinguish it from $\b$) that is expressed as $\ket{I} = \ket{{J}; \d}$ with $I\ne J$.
But the only temperature that dS space can sustain is the GH temperature which means that ${1}/{\b_{\rm dS}}=T_{\rm dS}=H/2\pi=1/\d$.
It is then sufficient to know the form of the thermal dS-scalar propagator for some generic temperature and then set $\b=\b_{\rm dS}$.

\section{The spectral index with thermal effects} \label{SpectralIndex}

The propagators in \eq{SKDininbeta} and \eq{Doutoutb} determine several important observables. 
At equal space-time points and at the time of horizon exit, defined as $|\tau | H =1$ and 
concentrating on horizon exiting modes specified by $|\tk \tau |\lesssim 1$,
they determine various cosmological indices derived from the thermal scalar power spectrum \cite{MolinaParis} (here ${\bf 1}$ is the $2\times 2$ matrix with unit elements)
\be\label{PSb}
P_{S,\b} {\bf 1}={\cal D}_\b {\bf 1} \vert_{\tau _1=\tau _2}\, ,
\ee
in terms of a single parameter (when the temperature takes its natural value $T=T_{\rm dS}$):
\be
\k \equiv \o_\tmk |\tau | \Bigl|_{|\tk \tau |=1} = \sqrt{\left( \tmk^2 + m^2_{\rm dS} \right) |\tau |^2}\Bigl|_{|\tk \tau |=1} =\sqrt{\frac{5 -d^2}{4}+M^2}\,
\ee
where $\omega_\tmk$ is the frequency defined under \eq{eom}. 
This parameter can be traded for the weight of the Hankel function,
as determined by the Klein-Gordon equation, in \eq{ncl}.
Of special importance in $d=3$ is the choice $M=0$, or $\k=i$, which is known to generate a scale invariant CMB spectrum.
This corresponds to $\n_{\rm cl}=\frac{3}{2}$ and decaying modes at the time of exit.

A particularly useful point of view \cite{AntMott} is to recognize the system at $\tau =-\infty$ as related to a UV Conformal Field Theory (CFT) labeled by the weight $\n_{\rm cl}$ and associated with
the Gaussian fixed point of the $d=3$ real scalar theory, that flows towards an interacting IR fixed point and the corresponding ${\rm CFT}$ at $\tau =0$.
It is clear that in the present context, exact scale invariance is realized in the $\ket{\rm in}$ vacuum, with the deviations
generated by a spontaneous shift in $M$ that, according to \eq{Doutoutb}, should have a finite temperature origin. 
Deviations can be encoded in general in a shift of the weight
$\n_{\rm cl}\to \n=\n_{\rm cl} + \n_{\rm q}$
that can be interpreted as a shift in the scaling dimension of a dS scalar field
\be
\D_{-} = \frac{d}{2} - \n = \frac{d}{2} - \n_{\rm cl} - \n_{\rm q} = \D_{\rm cl,-} - \n_{\rm q}\, .
\ee
There is a corresponding shadow partner solution to this with $\D_{+} = \frac{d}{2} + \n$. 
In this letter, we will be concerned with $(\D_-,\D_+)_{\rm cl}=(0,3)$.

In order to understand $\n_{\rm q}$ (which will turn out to be a non-trivial zero) we first point out that the 
$\ket{\rm out;\b}$ ($\b>\b_{\rm dS}$) state is a BT of the Bunch-Davies vacuum.
The mode functions before and after the transformation solve the same Bessel equation with frequency $\o_\tmk$.
Upon a time-dependent BT however, the frequency that an observer sees for a time other than his own, is \cite{Prokopec}:
\be
\O_\tmk = \o_\tmk (|c|^2+|s|^2)\, .
\ee
As a result, the horizon exit parameter is transformed as
\be
\k\to \L = \k \left(1 + 2 \frac{e^{-2 x\k} }{1-e^{-2x\k}}  \right) = \k \coth (x \k) \label{La}\, ,
\ee
where we have defined the dimensionless temperature parameter $x=\frac{\pi H}{2\pi T}$, that takes values in $[\pi, \infty]$. 
The transformed state in general has a reduced isometry with respect to the Bunch-Davies state.
This can be seen by the fact that the BT introduces a non-zero mass term $(\m_H^2+\xi\frac{{\cal R}}{H^2}) a^2 H^2\phi^2$ in the Lagrangian 
with exit parameter $\L^2 = |k\tau|^2 + a^2 \left[\m_H^2+(\xi-\frac{1}{6})\frac{\cal R}{H^2})\right]$ and that
the late time equations of motion 
\be\label{eomphi}
{\phi^{\prime\prime}} + 2aH{\phi^\prime} + \left(\m_H^2+\xi\frac{{\cal R}}{H^2}\right)a^2H^2\phi=0\, , \hskip .5cm {H^\prime} = -\frac{1}{2a}{\phi^\prime}^2
\ee
have no non-trivial solution with $H={\rm const.}$ and a non-zero, finite mass term.

The two limiting values of $x$ are interesting. Its natural value $x=\pi$ where $T=T_{\rm dS}$ 
gives $\L=\infty$ for $\k=i$. This is a special case where we recover a dS solution of maximal isometry
that corresponds to $\ket{\rm out;\b_{\rm dS}}$. As in the BD vacuum, no modes are seen to exit the horizon, this time due to their ultra-short wavelength.
In the limit $x\to \infty$ on the other hand, the out observer sees modes of any 
wavelength as exiting modes, since in this limit the time of exit approaches the horizon. This means that if he calls his frequencies $\O_\tmk$, 
then his horizon exit parameter will be forced to $\L_0\equiv \lim_{\tau \to 0}(\O_\tmk \tau) \to 0$.\footnote{In this limit $x$ becomes an odd multiple of $\pi/2$.}
This suggests to construct a trajectory from $(\L,x)\sim (\infty,\pi)$ to $(0,\infty)$ along which the value of some
yet to be defined thermal effect is kept non-zero and constant,
starting from a position a bit shifted away from the scale invariant limit $(\infty,\pi)$.
Deviations from exact dS isometry due to finite temperature effects can be encoded in the shift of the spectral index of scalar curvature fluctuations \cite{Bassett}
\be
n_{S,\b} = 1 + \frac{d \ln \left(\tmk^3 P_{S,\b}\right)}{d \ln \tmk}\,
\ee
where $P_{S,\b}$ is the thermal scalar power spectrum defined in \eq{PSb}.

In the previous section, we showed that the SK and TFD formalisms result to equivalent propagators. Consequently, from \eq{PSb} they both determine the same thermal deviation
\be\label{dnsx}
\d n_S \equiv n_{S,\b} -  1  = - \frac{2x}{\L} \left[ \frac{e^{-x\L} }{1-e^{-2x\L}}  \right]\, ,
\ee
of $n_S$ away from unity. Observe that in $\ket{\rm out;\b_{\rm dS}}$ where $x=\pi$ and $\L=\infty$, $\d n_S$ vanishes and we see a scale invariant spectrum.
Moving a bit away from it, $x\gtrsim \pi$,\footnote{It is implicitly assumed here that moving away from $T_{\rm dS}$ is a result
of spontaneous breaking of scale invariance, which is expected to lower the temperature.}
the state is $\ket{\rm out;\b}$ and $\d n_S$ becomes a one-parameter expression of $\L$. We can fix this freedom by determining the
value $n_{S,\b}$ by interpreting its deviation from unity as an anomalous dimension in the dual field theory in the spirit of the dS/CFT correspondence. 
Then we can reach $x=\infty$ along a trajectory which keeps this value constant for all temperatures.

In \cite{FotisAntonis1} it is proposed that within the dual field theory that lives on the horizon, the anomalous dimension
that shifts the spectral index is the critical exponent $\eta$, whose non-perturbative value is around $0.036$. Thus, near the horizon
\be \label{nSS}
n_S \simeq 1 - \eta = 0.964\,
\ee
while the experimentally measured value is equal to \cite{Planck2018X}: 
\bad
n_{S, \rm exp} = 0.9649 \pm 0.0042.
\ead
A known fact about the dual field theory of dS is that it is expected to be non-unitary \cite{Maldacena1}. 
Thus one could argue that the Ising model (or its large $N$ relatives) which is unitary, is not a good candidate as the dual to dS theory. 
The suggestion made here is that the field theory dual to the AdS version of the model under discussion 
-which could be in the universality class of the Ising model (or its large $N$ relatives)- is an analytic continuation of dS. 
This is complemented by the fact that \eq{nSS} is invariant under an analytic continuation (see \cite{FotisAntonis1} for a more detailed justification).

This is a constraining statement that leaves no free parameters.
In \cite{FotisAntonis1} it is shown that the quantity by which $\D_{+,{\rm cl}}$ shifts is the operator anomalous dimension of the
trace of the Ising stress energy tensor $\Theta$, which is an exact zero, realized as the cancellation 
$\G_\Theta=\gamma_{\cal O}-2 \gamma_\sigma$ = 0, where $\gamma_{\cal O}$ is the ``total'' operator anomalous dimension
and $\gamma_\sigma$ is the field anomalous dimension, or the so called wave function renormalization.
It is therefore in this sense that $\n_q$ is a non-trivial zero, being related to the vanishing anomalous dimension of a special operator that is the energy-momentum tensor.
In \cite{FotisAntonis1} it is also demonstrated that it is the total anomalous dimension $\gamma_{\cal O}$ that ends up shifting the spectral index $n_S$.  
Of course, outside the fixed point where the Ising field is massive, $M$ deviates from zero in the bulk and the solution to \eq{eomphi} is not dS. 
It is important however to understand that the main effect on $n_S$ comes from the critical value $\eta$ of $2 \gamma_\sigma$ and the
deviation from the critical value is small as long as the system sits in the vicinity of the fixed point.
For this reason the leading order results are independent of the source of the breaking. 
In a sense the only assumption here is that there is a mechanism of spontaneous breaking of scale invariance.
From the point of view of the boundary this could be for example justified as some sort of a Coleman-Weinberg mechanism.

\section{Line of constant physics and other observables}

A line of constant physics (LCP) is a set of points on the phase space upon which the value of a physical quantity remains fixed. 
What we will demonstrate now is that in the bulk, there is a LCP, labelled by the fixed value $\d n_S =-\eta$, along which the system is 
heated up from zero temperature where $\L_0= 0$ and $x= \infty$, up to the dS temperature.  A few points on this line and a picture of the LCP can be found in Fig. 1.
%
\begin{figure}[!t]
\begin{minipage}{9cm}\label{figLCP}
\begin{center}
\begin{tabular}{|c|c|}
\hline 
$\L$ & $x$ \\
\hline \hline   
$\to 0$  & $\to \infty$ \\ \hline
$10^{-6}$  & $3.5\cdot 10^{7}$ \\ \hline
$0.01$  & $1600$ \\ \hline
$0.5$  & $14.8$ \\ \hline
$\to 1.5117$  & $\to\pi$ \\ \hline
\end{tabular}
\end{center}
\end{minipage}
\begin{minipage}{9cm}
\includegraphics[width=5cm]{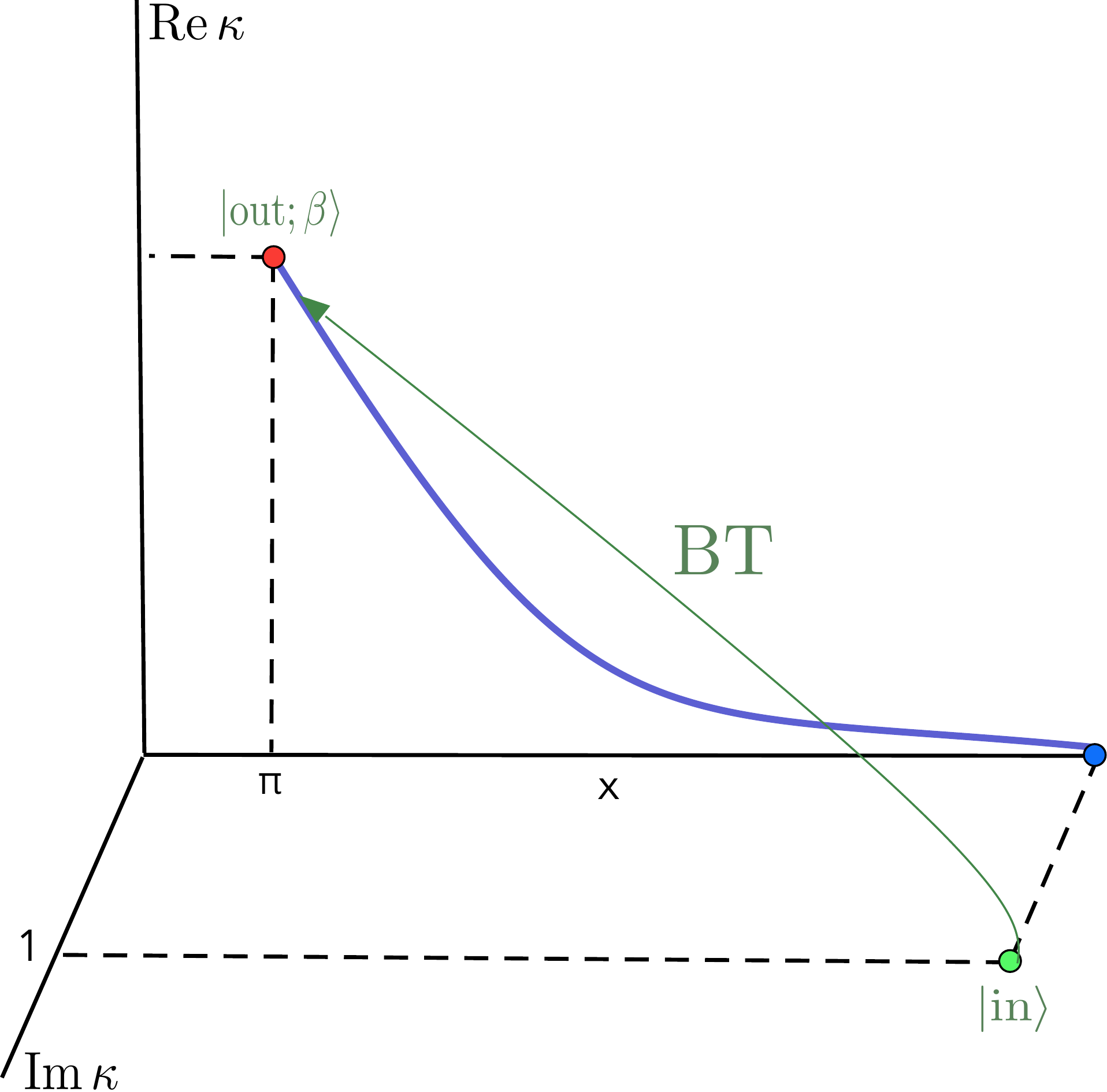}\hfill
\end{minipage}
\caption{
Left: A few points of the nearly conformal LCP defined by $\d n_S=-\eta$.
Right: The Bogolyubov Transformation $\ket{\rm in}\to\ket{\rm out; \b}$ and the LCP, on the complex plane where $\k=\L+i\,{\rm Im \k}$.} \label{LCP2}
\end{figure}
%
We stress that for a given $x$ the corresponding value of $\L$ is fixed by the label of the LCP.
Thus near the endpoint of the LCP where $x\simeq\pi$, the value $\L_\pi\simeq 1.5117$ is a fixed output. 
It is important to emphasize that the LCP is really meaningful up to just outside its two limiting points.
Up to around $x\simeq \pi$ it is characterized by a non-zero $\d n_S$ which however at exactly $x=\pi$ becomes equal to zero, since 
the trace of the boundary stress-energy tensor to which the bulk scalar couples, vanishes.
Analogously, the interpretation of each point on it as a dS space of the same $T_{\rm dS}$ is possible
everywhere except at $x=\infty$, where the intrinsic temperature must become abruptly unobservable.


Since there are no free parameters, several other observables that are determined by $P_{S,\beta}$ are expected to be also fixed.
Define for example the moment
\be
n_{S,\b}^{(1)} = \frac{d n_{S,\b}}{d \ln \tmk}\, 
\ee
and let us compute it using that $n_{S}^{(1)}=0$. The result, evaluated under the same conditions as $n_{S,\b}$, is 
\be
n_{S,\b}^{(1)} = \d n_S \left[ 2 - \frac {1}{\L^2} - \frac{x}{\L} \left(1 + \frac{2e^{-2x\L}}{1-e^{-2x\L}}\right)\right]
\ee
which, substituting $x\simeq\pi$ and $\L=\L_\pi\simeq1.5117$, gives
\be
n_{S,\b}^{(1)} = 0.0186
\ee
for the running of the index. The constraints given in \cite{Planck2018X} are:
\bad
n_{S,\text{exp}}^{(1)} = 0.013 \pm 0.012.
\ead

Finally, the universal contribution to the non-Gaussianity parameter \cite{Creminelli}, can be expressed 
in terms of $N= \int^{t_f}_{t_i} dt H $ and its derivatives in the in-vacuum, as \cite{KR1}
\be
f^{\rm un}_{NL} = \frac{5}{6} \frac{N_{\r\r}}{N_\r^2}
\ee
with $N_\r= \frac{\partial N}{\partial \r}$, $N_{\r\r}= \frac{\partial^2 N}{\partial \r^2}$ and 
$\r\equiv P_{S,\b}$.\footnote{The parameter $f_{NL}$ is defined by a more general expression \cite{Maldacena1}. The expression we use is given in \cite{KR1} for the special case of a single scalar field.} It is computed to be
\bad
f^{\rm un}_{NL}  = -\frac{5\Bigl[ x ( -1 + \Lambda^2 )^2 \Bigl(1  + x \Lambda  \cot(\frac{x \Lambda}{2}) \Bigr) + 
2 \Lambda^3 \sinh (x \Lambda) \Bigr]}{6 \Lambda^2 \Bigl[ x (-1 + \Lambda^2) + \Lambda \sinh (x \Lambda)\Bigr]}\, .
\ead
For $x\simeq\pi$ and $\L=\L_\pi\simeq1.5117$ this gives
\bad \label{fNLbeta}
f^{\rm un}_{NL} = -1.7138
\ead
while one of the experimental results for $f^{\rm un}_{NL}$ in a certain analysis is \cite{Planck2018IX}: 
\bad
f^{\rm un}_{NL,\text{exp}} = - 1.7 \pm 5.2.
\ead 
\section{Conclusion}

We considered a thermal scalar in de Sitter background.
Starting from the Bunch-Davies $\ket{\rm in}$ vacuum, a Bogolyubov Transformation placed us in the interior of the finite temperature phase diagram
in a thermal state $\ket{\rm out;\b}$.
This state can be connected through holography to the vicinity of an interacting IR fixed point, in the universality class of the 3d Ising model.
The system in this state is rather special, in the sense that the boundary operator that couples to the scalar curvature perturbations in the bulk
has a classical scaling dimension.
The critical exponent $\eta$ is the order parameter of the breaking of the scale invariant spectrum of curvature fluctuations and 
a simple argument from the dS/CFT correspondence fixes the parametric freedom in the dS scalar theory,
yielding the prediction $n_S=0.964$.
We also computed in the same context additional cosmological observables
such as the first moment of the scalar spectral index and the non-Gaussianity bispectrum parameter $f_{NL}$ and evaluated them numerically.
Our predicted values of $n_S$, $n_{S,\b}^{(1)} $ and $f_{NL}$ are well within current experimental bounds \cite{Planck2018X, Planck2018IX}.

\vskip .1cm
{\bf Acknowledgments.}
The research of F.K. leading to these results has received funding from the Norwegian Financial Mechanism for years 2014-2021, grant nr DEC-2019/34/H/ST2/00707.
The authors thank I. Dalianis for discussions.

\begin{appendix}
\section{Appendix}\label{AppKMS}

In this Appendix we discuss the real time construction in the Hamiltonian formulation.
First we give a shortcut derivation of the thermal propagator \eq{SKDininbeta} that starts from flat space and the definitions
\bea
{\cal D}^\beta_{+-}(\tau_1,\tau_2) &=& W_2(\tau_1,\tau_2) + W_1(\tau_1,\tau_2) \nl
{\cal D}^\beta_{-+}(\tau_1,\tau_2) &=& W_1(\tau_2,\tau_1) + W_2(\tau_2,\tau_1) 
\eea
with the Wightman functions defined as
\bea\label{Wightmans}
&& W_1(\tau_1,\tau_2) \equiv \frac{{\rm Tr} \{ a^\dagger(\tau_1) a(\tau_2)\, \rho \}}{{\rm Tr} \{ \rho \}} = n_B e^{i\omega (\tau_1-\tau_2)}\nl
&& W_2(\tau_1,\tau_2) \equiv \frac{{\rm Tr} \{ a(\tau_1) a^\dagger(\tau_2)\, \rho \}}{{\rm Tr} \{ \rho \}} = (1+n_B) e^{-i\omega (\tau_1-\tau_2)}
\eea
where $\rho = e^{- \beta {\cal H}}$ is the thermal density matrix, ${\cal H}$ is the (harmonic oscillator) Hamiltonian and the second equalities show the result of the trace computations. 
Now since the time dependent part of the mode function in flat space is $u(\tau) = e^{i\omega\tau}$ we can write $e^{i\omega (\tau_1-\tau_2)} = u(\tau_1) u^*(\tau_2)$
and pass to dS space via the substitution $u(\tau)\to \chi_\tmk(\tau)$. Then indeed
\bea
{\cal D}^\beta_{+-}(\tau_1,\tau_2) &=& \chi^*_\tmk(\tau_1) \chi_\tmk(\tau_2) + n_B(\beta) \left( \chi_\tmk(\tau_1)\chi_\tmk^*(\tau_2) + \chi_\tmk^*(\tau_1)\chi_\tmk(\tau_2)\right)\nl
{\cal D}^\beta_{-+}(\tau_1,\tau_2) &=& \chi_\tmk(\tau_1) \chi^*_\tmk(\tau_2) + n_B(\beta) \left( \chi_\tmk(\tau_1)\chi_\tmk^*(\tau_2) + \chi_\tmk^*(\tau_1)\chi_\tmk(\tau_2)\right)
\eea
and by imposing ${\cal D}^\beta_{++}(\tau _1;\tau _2)  = \theta( \tau _1 - \tau _2 ) {\cal D}^\beta_{-+}(\tau _1; \tau _2) +  \theta( \tau _2 - \tau _1 ) {\cal D}^\beta_{+-}(\tau _1; \tau _2) $,
${\cal D}^\beta_{--}(\tau _1; \tau _2) = {\cal D}^{^\beta*}_{++}(\tau _1; \tau _2)$ and applying for $\beta/2$, we arrive again at \eq{SKDininbeta}.

A thermal propagator has to satisfy a variant of the KMS condition. 
The KMS condition originates from the definition
\bad \label{KMS}
\braket{\phi(t_1,x_1) \phi(t_2,x_2)}_\b = \frac{{\rm Tr} \left\{ \phi(t_1,x_1) \phi(t_2,x_2) \rho \right\}}{{\rm Tr} \{\rho \}} 
\ead
that leads, in principle, to the thermally corrected dS propagator. However for time dependent Hamiltonians the direct computation of the trace is not obvious. 

The condition takes a simple form though near $\tau\to -\infty$, which we can show explicitly.
In Thermofield Dynamics, the form of the KMS condition depends on a gauge parametrized by a real number, say $\a$. 
It is a well known fact that TFD propagators satisfy such a condition in any of these $\a$-gauges \cite{Henning, Ojima}. 
The condition holds due to the relation \cite{Takahashi}
\bad \label{a=1/2}
a^-_\tk \ket{0;\beta} = e^{- \a\b \o_\tmk} \tilde a^+_\tk \ket{0;\beta}, \qquad \bra{0;\b} a^+_\tk = \bra{0;\b} \tilde a^-_\tk e^{- (1-\a) \b \o_\tmk}
\ead   
between the standard annihilation operator acting on the vacuum of the Hilbert space ${\cal H}$ and the tilded creation operator acting on the vacuum of ${\tilde{\cal H}}$. 
The thermal vacuum $\ket{0;\beta}$ is defined by the action of a unitary operator on the tensor product of the vacuum states in ${\cal H} \times {\tilde{\cal H}}$.
Using the above relations, one can straightforwardly show that the Wightman function 
between the fields $\Phi, \tilde \Phi$ for $\a=1/2$ satisfies the condition (the mode functions of the scalar field near $\tau \rightarrow - \infty$ reduce to plane waves):
\bad
\braket{0;\b | \Phi(\tau_1, \tx) \tilde \Phi(\tau_2,\ty) | 0;\b} = \braket{0;\b | \Phi(\tau_2 + i \frac{\b}{2},\ty) \tilde  \Phi(\tau_1 - i \frac{\b}{2}, \tx)  | 0;\b} ,
\ead 
which is the KMS condition in the $\a=1/2$ gauge. This is a relevant for us case, since the transformation matrix \eq{unitaryDas} is in this gauge \cite{Henning}.
A different gauge choice is to take $\a=1$, where
\bad
a^-_\tk \ket{0;\b} = e^{-\b \o_\tmk} \tilde a^+_\tk \ket{0;\b}, \qquad \bra{0;\b} a^+_\tk = \bra{0;\b} \tilde a^-_\tk.
\ead 
Then, the same Wightman function as above needs to satisfy 
\bad
\braket{0;\b | \Phi(\tau_1, \tx) \tilde \Phi(\tau_2,\ty) | 0;\b} = \braket{0;\b | \Phi(\tau_2,\ty) \tilde  \Phi(\tau_1 - i \b, \tx)  | 0;\b},
\ead
the KMS condition in the $\a=1$ gauge. Note that this is a relation where the usual form of the KMS condition of thermal field theory can be recognised.
These two gauges however can be readily seen to correspond to equivalent Wightman functions,
as they can be related by a shift in the imaginary time, by $\tau_{1,2}\longrightarrow \tau_{1,2}-i\frac{\b}{2}$.
By this shift freedom, one can also see that the diagonal elements of the propagator do not satisfy any non-trivial constraint.
In conclusion, to the extent that \eq{KMS} applies to dS space and the trace is computable, 
the thermal propagator it defines satisfies a KMS condition.

\end{appendix}



\end{document}